\documentclass[aps,prl,reprint,amsmath,amssymb]{revtex4-2}

\usepackage{graphicx}
\usepackage{dcolumn}
\usepackage{bm}

\begin{document}

\preprint{APS/123-QED}

\title{Impact of Helicity–Isospin Convention Matching on the Extraction of $\boldsymbol{\alpha}$ from $\boldsymbol{B^0\to(\rho\pi)^0}$}

\author{Radek \v{Z}leb\v{c}\'{i}k}
 \email[Contact author: ]{radek.zlebcik@matfyz.cuni.cz}
 \affiliation{Institute of Particle and Nuclear Physics, Faculty of Mathematics and Physics, Charles University, Prague, Czech Republic}

\date{\today}

\begin{abstract}

We show that the $\rho\to\pi\pi$ helicity convention used in the time-dependent Dalitz-plot measurements of $B^0\to(\rho\pi)^0$ decays does not match the convention assumed in the subsequent isospin analysis. Once corrected, the tension between the direct determination of the unitarity-triangle angle $\alpha$ from these decays and its indirect determination from global fits decreases from $2.9\sigma$ to $2.1\sigma$. Combined with the $B\to\pi\pi$ and $B\to\rho\rho$ modes, the direct determination becomes $\alpha = (87.0^{+11.8}_{-4.7})^\circ$, compared with the original world average $(84.1^{+3.7}_{-3.0})^\circ$.

\end{abstract}

\maketitle

In the Standard Model, \textit{CP} violation originates from a single
irreducible phase of the unitary Cabibbo-Kobayashi-Maskawa (CKM)
matrix~\cite{Cabibbo:1963yz,Kobayashi:1973fv}.
This picture is tested by overconstraining the sides and angles of the unitarity
triangle, which are determined redundantly from a large set of flavor
observables~\cite{Charles:2004jd,UTfit:2022hsi}.
The angle $\alpha = \arg(-V_{td}V^{*}_{tb} / V_{ud}V^{*}_{ub})$ is the least
precisely determined of the three angles, and it is also the only one for which the
direct measurement and the indirect, unitarity-based determination differ by
about $2\sigma$~\cite{HeavyFlavorAveragingGroupHFLAV:2024ctg,Charles:2004jd}.

Direct access to $\alpha$ is provided by isospin
analyses~\cite{Gronau:1990ka,Lipkin:1991st} of the $B\to\pi\pi$,
$B\to\rho\rho$, and $B^0\to(\rho\pi)^0$ systems, which disentangle the tree and
penguin contributions using the full set of charge combinations of each
mode~\cite{HeavyFlavorAveragingGroupHFLAV:2024ctg}.
The values extracted from $B\to\pi\pi$ and $B\to\rho\rho$ are compatible with
the indirect determination; the one extracted from $B^0\to(\rho\pi)^0$ is not.
Combining the Belle~\cite{Belle:2007krm,Belle:2007jkw} and
BaBar~\cite{BaBar:2013uwm} time-dependent Dalitz measurements, the $1\sigma$
interval from $B^0\to(\rho\pi)^0$ alone is
$[43.8^\circ, 61.8^\circ] \cup [136.4^\circ, 146.6^\circ]$, as found in Ref.~\cite{Charles:2017evz},
far from the indirect value, with which it is in tension at the level of 
$2.9\sigma$~\cite{Charles:2017evz,HeavyFlavorAveragingGroupHFLAV:2024ctg}.
Because the corresponding $\chi^2$ profile is steep, this single mode carries a
large weight in the world average and pulls the combined HFLAV value down to
$\alpha_\mathrm{HFLAV} = (84.1^{+3.7}_{-3.0})^\circ$~%
\cite{HeavyFlavorAveragingGroupHFLAV:2024ctg}, to be compared with
$(91.2^{+2.0}_{-1.0})^\circ$ obtained by the CKMfitter group from a global fit
in which the modes directly sensitive to $\alpha$ are removed~\cite{Charles:2004jd}.
The $B^0\to(\rho\pi)^0$ channel is thus the origin of the apparent
direct--indirect discrepancy rather than an independent confirmation of it.

In this Letter, we show that this deviation is to a large extent caused by a
mismatch between the $\rho\to\pi\pi$ helicity convention used in the
$B^0\to(\rho\pi)^0$ measurements and the convention assumed in the subsequent
isospin analysis.
We derive the angular dependence of the $\rho\to\pi\pi$
decay explicitly, which the original proposal of Snyder and
Quinn~\cite{Snyder:1993mx} does not provide, identify the helicity convention
implied by the isospin relation, and repeat the $\alpha$ determination with the
existing data expressed in a consistent convention.

In 1993, \textit{Snyder and Quinn} \cite{Snyder:1993mx} proposed an experimental method that uses interference regions in the Dalitz time-dependent analysis of the $B^0\to \pi^+\pi^-\pi^0$ decay to fully constrain $B^0\to\rho^+\pi^-$, $B^0\to\rho^-\pi^+$, and $B^0\to\rho^0\pi^0$ two-body amplitudes.
Exploiting the isospin relation introduced in Ref.~\cite{Lipkin:1991st}, in principle, it allows $\alpha$ measurement without ambiguities.
The three-body $B^0\to \pi^+\pi^-\pi^0$ decay amplitudes can be written as
\begin{equation}
\begin{split}
   A_{3\pi} = f_+(m_{+0})\cos \theta_+\, A^+ + f_-(m_{-0}) \cos \theta_-\, A^- \\
     +\, f_0(m_{+-}) \cos \theta_0\, A^0,\\
   \bar{A}_{3\pi} = f_+(m_{+0})\cos \theta_+\, \bar{A}^+ + f_-(m_{-0}) \cos \theta_-\, \bar{A}^- \\
     +\, f_0(m_{+-}) \cos \theta_0\, \bar{A}^0,
\label{eq:dalitz_amplitude}
\end{split}
\end{equation}
where $A^i$ or $\bar{A}^i$ are the amplitudes of the $B^0\to \rho^i \pi^{-i}$ or $\bar{B}^0\to \rho^i\pi^{-i}$ transitions, $m_{kl}$ is the invariant mass of $\pi^k\pi^l$ system, and $f_i(m_{kl})$ are the $\rho$ meson lineshapes (Breit-Wigner in the original Snyder-Quinn~paper)~\footnote{In the Belle and BaBar measurements more advanced Gounaris-Sakurai $\rho$ lineshapes are used and Belle uses Blatt-Weisskopf form factors. These differences are irrelevant for the following discussion.}.
The $\bar{A}^i$ and $\bar{A}_{3\pi}$ amplitudes for $\bar{B}^0$ are, for simplicity, scaled by $B^0\!-\!\bar{B}^0$ mixing factor $q/p$.
The angular factors $\cos \theta_i$ correspond to the helicity angles.
These are evaluated in the $\rho^i$ rest frame, where the spectator pion flies in $-z$ direction.
Depending on the convention, the angle of the first or second pion in the $\rho \to \pi \pi$ can be considered.
In the original Snyder-Quinn paper or in the follow-up publication, we have not found any discussion of which of the two pions in the $\rho$ decay should be used as a reference daughter for the helicity angle definition for $\rho^+$, $\rho^-$, and $\rho^0$.
Swapping the reference daughter pion in the helicity angle definition leads to a sign flip of the angular factor $\cos \theta_i$ or, equivalently, to a sign flip of the related amplitude $A^i$ in the Dalitz $A_{3\pi}$ amplitude of Eq.~(\ref{eq:dalitz_amplitude}).

\textit{Angular dependence of the $\rho\to\pi\pi$ decay}---We now derive the
angular factors $\cos\theta_i$ entering Eq.~(\ref{eq:dalitz_amplitude}) in two
sign conventions for the $\pi$ and $\rho$ isospin states, which makes explicit
which quantities are convention dependent.

Lipkin-Nir-Quinn-Snyder convention:\quad The pentagon isospin relation for $B\to\rho\pi$ was introduced by Lipkin et al.~\cite{Lipkin:1991st}, and it was later adapted by Snyder and Quinn~\cite{Snyder:1993mx} in the proposed $B^0\to \pi^+\pi^-\pi^0$ analysis that uses interference regions for constraining relative phases between $B^0\to \rho^i\pi^{-i}$ amplitudes.

The $\pi\pi$ and $\rho\pi$ isospin decompositions in Eqs.~(4) and (28) of Ref.~\cite{Lipkin:1991st} imply that the $\pi$ and $\rho$ states obey the convention
\begin{equation}
   |1, +1\rangle =  |\pi^+\rangle, \quad
   |1,\:\:\: 0\rangle  =  |\pi^0\,\rangle,\quad
   |1, -1\rangle =  |\pi^-\rangle,
   \label{eq:sqConventionSQ}
\end{equation}
up to an overall irrelevant phase.

In this sign convention, the pentagon isospin relation for the $B\to \rho\pi$ system has the following shape \cite{Lipkin:1991st}
\begin{eqnarray}
   A^+ + A^- + 2A^0                   = \sqrt{2}\, (A^{+0} + A^{0+}),\nonumber\\
   \bar{A}^+ + \bar{A}^- + 2\bar{A}^0 = \sqrt{2}\, (A^{-0} + A^{0-}),
   \label{eq:isospinRelation}
\end{eqnarray}
where $A^i$ and $\bar{A}^i$ amplitudes for neutral $B$ are introduced in~Eq.~(\ref{eq:dalitz_amplitude}).
The $A^{i0}$ and $A^{0i}$ are amplitudes of $B^i\to \rho^i \pi^0$ and $B^i\to \rho^0 \pi^i$ decays involving charged $B$ meson.
The $A^{-0}$ and $A^{0-}$ amplitudes incorporate $q/p$ conventional factor, similarly to $\bar{A}^i$ amplitudes.

It can be shown that the $A = A^+ + A^- + 2A^0$ linear combination of the $B^0\to (\rho\pi)^0$ amplitudes proceeds only through $\Delta I=3/2$ transition and is free from the QCD penguin contributions and, consequently, in the isospin limit and neglecting electroweak penguin contributions, the CKM angle $\alpha$ can be calculated as~\cite{Lipkin:1991st}
\begin{equation}
   \mathrm{e}^{2 i \alpha} = \frac{\bar{A}^+ + \bar{A}^- + 2\bar{A}^0}{A^+ + A^- + 2A^0}.
   \label{eq:alpha}
\end{equation}

The total wave function of the two-pion state from $\rho$ decay has to be symmetric, and since the angular part is in an antisymmetric $l=1$ state, the isospin part is also an antisymmetric $I=1$ isospin state.
It reads as follows
\begin{eqnarray}
   |+\rangle_\rho &=& \frac{1}{\sqrt{2}} \left( |+\rangle_\pi|0\rangle_\pi - |0\rangle_\pi|+\rangle_\pi \right),\nonumber\\
   |0\rangle_\rho &=& \frac{1}{\sqrt{2}} \left( |+\rangle_\pi|-\rangle_\pi - |-\rangle_\pi|+\rangle_\pi \right),\nonumber\\
   |-\rangle_\rho &=& \frac{1}{\sqrt{2}} \left( |0\rangle_\pi|-\rangle_\pi - |-\rangle_\pi|0\rangle_\pi \right),
   \label{eq:rhoDecaysGen}
\end{eqnarray}
where, for simplicity, $|1, i\rangle$ is labeled as $|i\rangle$.
Using the convention introduced in Eq.~(\ref{eq:sqConventionSQ}) this leads to the following isospin wave functions of the $\rho$ mesons
\begin{eqnarray}
   |\rho^+\rangle &=& \frac{1}{\sqrt{2}} \left( |\pi^+\pi^0\rangle - |\pi^0\pi^+\rangle \right),\nonumber\\
   |\rho^0\rangle &=& \frac{1}{\sqrt{2}} \left( |\pi^+\pi^-\rangle - |\pi^-\pi^+\rangle \right),\nonumber\\
   |\rho^-\rangle &=& \frac{1}{\sqrt{2}} \left( |\pi^0\pi^-\rangle - |\pi^-\pi^0\rangle \right).
   \label{eq:rhoDecays}
\end{eqnarray}
The scalar nature of the $B$ in $B^0\to\rho^i\pi^{-i}$ implies that the orbital wave function is $Y^{m=0}_{l=1}(\theta,\varphi) \propto \cos \theta$.
Based on the ordering in Eq.~(\ref{eq:rhoDecays}), one gets the following assignment of the angles entering Eq.~(\ref{eq:dalitz_amplitude})
\begin{alignat}{2}
  \rho^+ &\to \pi^+\pi^0 &&: \cos\theta_{\pi^+}, \nonumber\\
  \rho^0\, &\to \pi^+\pi^- &&: \cos\theta_{\pi^+}, \nonumber\\
  \rho^- &\to \pi^-\pi^0 &&: \cos\theta_{\pi^0}.
  \label{eq:helicityConvention}
\end{alignat}
In this convention, the particle with the higher charge is taken as the reference one.

Beneke-Neubert convention:\quad  For the sake of clarity, we also show the derivation of the angular spectra in an alternative sign convention used, e.g., by Beneke and Neubert~\cite{Beneke:2003zv}.

Beneke and Neubert relate both the $\pi$ and $\rho$ mesons with the following quark contents $\pi^+\sim \bar{d}u$, $\pi^0\sim \frac{1}{\sqrt{2}}(\bar{u}u - \bar{d}d)$, $\pi^-\sim \bar{u}d$, as given in Eq.~(5) of Ref.~\cite{Beneke:2003zv}.
Since the quark and antiquark isospin doublets are $(u, d)$ and $(\bar{d}, -\bar{u})$, the convention (up to overall phase) in Ref.~\cite{Beneke:2003zv} is
\begin{equation}
   |1, +1\rangle =  -|\pi^+\rangle, \quad
   |1,\:\:\: 0\rangle  =  |\pi^0\,\rangle,\quad
   |1, -1\rangle =  |\pi^-\rangle.
   \label{eq:sqConventionBN}
\end{equation}

The isospin relation in this convention, Eq.~(129) of Ref.~\cite{Beneke:2003zv}, has a flipped sign in front of the $A^0$ amplitude:
\begin{eqnarray}
   A^+ + A^- - 2A^0                   = \sqrt{2}\, (A^{+0} + A^{0+}),\nonumber\\
   \bar{A}^+ + \bar{A}^- - 2\bar{A}^0 = \sqrt{2}\, (A^{-0} + A^{0-}).
   \label{eq:isospinRelationBN}
\end{eqnarray}
The CKM angle $\alpha$ is then calculated as
\begin{equation}
   \mathrm{e}^{2 i \alpha} = \frac{\bar{A}^+ + \bar{A}^- - 2\bar{A}^0}{A^+ + A^- - 2A^0}.
   \label{eq:alphaBN}
\end{equation}

The isospin part of the wave function for the $\rho$ mesons is according to Eqs.~(\ref{eq:rhoDecaysGen}) and (\ref{eq:sqConventionBN})
\begin{eqnarray}
   |\rho^+\rangle &=& \frac{1}{\sqrt{2}} \left( |\pi^+\pi^0\rangle - |\pi^0\pi^+\rangle \right),\nonumber\\
   |\rho^0\rangle &=& \frac{1}{\sqrt{2}} \left( |\pi^-\pi^+\rangle - |\pi^+\pi^-\rangle \right),\nonumber\\
   |\rho^-\rangle &=& \frac{1}{\sqrt{2}} \left( |\pi^0\pi^-\rangle - |\pi^-\pi^0\rangle \right).
   \label{eq:rhoDecaysBN}
\end{eqnarray}
Repeating the same procedure as in the previous part, the following assignments for helicity angles are derived
\begin{alignat}{2}
  \rho^+ &\to \pi^+\pi^0 &&: \cos\theta_{\pi^+}, \nonumber\\
  \rho^0\, &\to \pi^+\pi^- &&: \cos\theta_{\pi^-}, \nonumber\\
  \rho^- &\to \pi^-\pi^0 &&: \cos\theta_{\pi^0}.
  \label{eq:helicityConventionCyclic}
\end{alignat}
The obtained result is known as the cyclic helicity convention.
It differs from the previous convention of Eq.~(\ref{eq:helicityConvention}) in the $\rho^0\to\pi^+\pi^-$ mode; therefore, the change of the convention effectively flips the $\cos \theta_0$ sign.

When the $A^+$, $A^-$, $A^0$ amplitudes are measured from the angular spectra of $B^0\to\pi^+\pi^-\pi^0$ decay using Eq.~(\ref{eq:dalitz_amplitude}), the sign of the fitted $A^0$ amplitude changes between the two discussed helicity conventions.
However, this sign flip is compensated by the different sign in the expression for $\alpha$, Eq.~(\ref{eq:alpha}) vs Eq.~(\ref{eq:alphaBN}), so that the measured value of~$\alpha$ is convention-independent.

The Belle experiment uses the cyclic convention for helicities, see Eq.~(19) of Ref.~ \cite{Belle:2007jkw}.
The BaBar experiment also uses the cyclic convention, as described below Eq.~(1) of Ref.~\cite{BaBar:2013uwm}.
The conventions of the experiments only differ by an irrelevant global sign flip of all three $\cos \theta_i$ components.
Therefore, both collaborations use the cyclic helicity convention of Eq.~(\ref{eq:helicityConventionCyclic}) while calculating~$\alpha$ using Eq.~(\ref{eq:alpha}), which holds in the convention of Eq.~(\ref{eq:helicityConvention}).

\begin{figure*}[t]
\centering
\begin{minipage}[t]{0.45\textwidth}
  \centering
  (a)\\[0pt]
  \includegraphics[width=\linewidth]{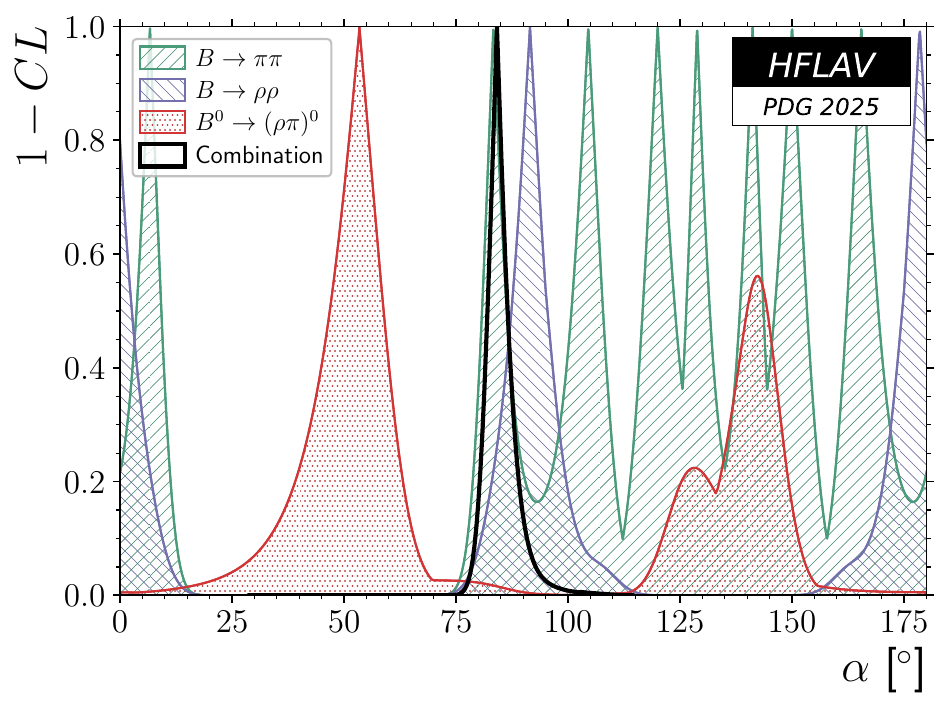}
\end{minipage}%
\hspace{0.03\textwidth}%
\begin{minipage}[t]{0.45\textwidth}
  \centering
  (b)\\[0pt]
  \includegraphics[width=\linewidth]{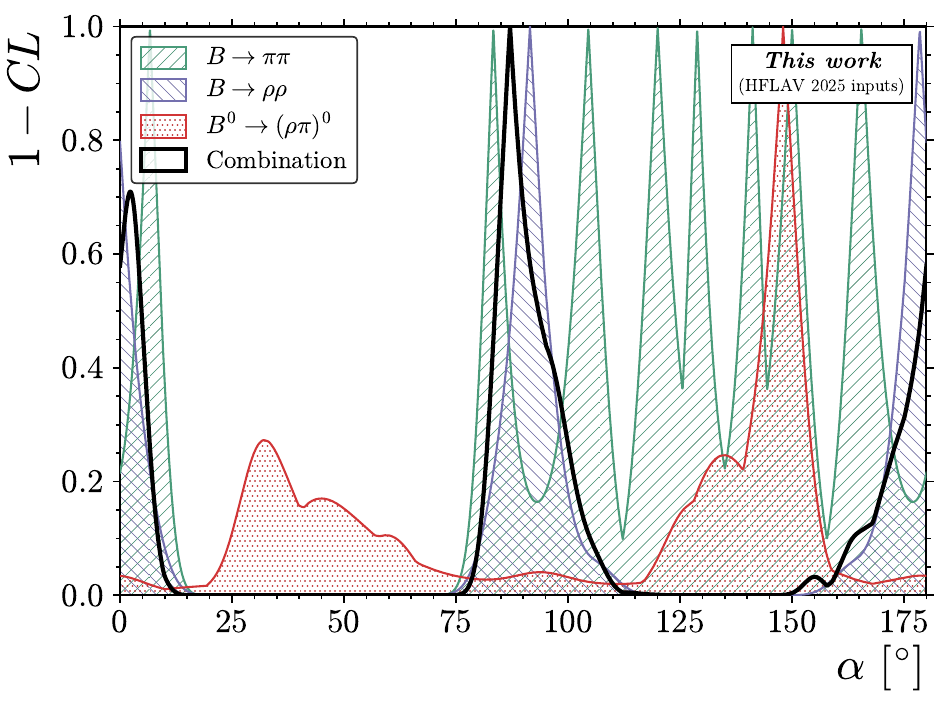}
\end{minipage}
\caption{One-dimensional scans of $\alpha$ in the $[0^\circ,180^\circ]$ range.
Panel (a) shows the results published by
HFLAV~\cite{HeavyFlavorAveragingGroupHFLAV:2024ctg}; panel (b) includes the
$B^0\to(\rho\pi)^0$ mode with a consistent helicity convention for
$\rho\to\pi\pi$ in the measurements and in the subsequent isospin analysis.
The $B\to\pi\pi$ and $B\to\rho\rho$ curves and all experimental inputs are
identical in both panels; only the $B^0\to(\rho\pi)^0$ curve and, consequently,
the combined curve differ.}
\label{fig:alphaScan}
\end{figure*}

\textit{Implications for the $\alpha$ determination}---The latest determination of the CKM angle $\alpha$ from the HFLAV group from 2025~\cite{HeavyFlavorAveragingGroupHFLAV:2024ctg} uses a frequentist approach to combine $B\to \pi\pi$, $B\to \rho\rho$, and $B^0\to (\rho\pi)^0$ channels.
The average is dominated by $B\to \rho\rho$ channel; however, the $B^0\to (\rho\pi)^0$ mode pushes the combined value slightly down, as can be seen from Fig.~\ref{fig:alphaScan}(a).
The extracted combined value is $\alpha_\mathrm{HFLAV} = (84.1^{+3.7}_{-3.0})^\circ$.

The time-dependent Dalitz analysis of $B^0\to (\rho\pi)^0$ was performed by BaBar~\cite{BaBar:2013uwm} and Belle~\cite{Belle:2007krm,Belle:2007jkw}.
Both collaborations use the cyclic helicity convention of Eq.~(\ref{eq:helicityConventionCyclic}) for the $\rho \to \pi \pi$ subdecay.
However, as shown above, when $\alpha$ is calculated by Eq.~(\ref{eq:alpha}), as both collaborations do, the helicity angle convention in Eq.~(\ref{eq:helicityConvention}) has to be used.
Since the only difference between these two helicity conventions is the sign of the $\cos \theta_0$ term, the measured amplitudes can be transformed to the Lipkin-Nir-Quinn-Snyder helicity convention by flipping the signs of $A^0$ and $\bar{A}^0$, see~Eq.~(\ref{eq:dalitz_amplitude}).
More precisely, both collaborations measure 26 $U/I$ parameters which are bilinear functions of $A^i$ and $\bar{A}^i$ and are defined, for example, in Ref.~\cite{Belle:2007jkw}.
These can be converted to the appropriate helicity convention by flipping the sign of the $U/I$ parameters where the $A^0$ or $\bar{A}^0$ amplitude enters linearly.
This is the case for the following 12 parameters: $U^{\pm,\mathrm{Re}}_{\pm0}$, $U^{\pm,\mathrm{Im}}_{\pm0}$, $I^{\mathrm{Re}}_{\pm0}$, $I^{\mathrm{Im}}_{\pm0}$.
In addition, the sign of off-diagonal elements of the covariance matrix has to be flipped accordingly.

First, as a cross-check, we repeated the $\alpha$ scan for the unchanged combined $U/I$ parameters \cite{HeavyFlavorAveragingGroupHFLAV:2024ctg}, where the combination has been performed by HFLAV based on $26 \times 26$ covariance matrices provided by both experiments.
In this way, we exactly reproduced the public HFLAV $B^0\to(\rho\pi)^0$ curve in Fig.~\ref{fig:alphaScan}(a).
In addition, for this scenario, we exactly reproduced the combined curve and combined $\alpha$ value, confirming that both the $\alpha$ fit and the combination procedure are consistent with HFLAV.

In the next step, the isospin $B^0\to(\rho\pi)^0$ analysis using the $U/I$ parameters expressed in the matched convention was repeated~\footnote{We verified that identical results are obtained when the $U/I$ parameters are left unchanged and $\alpha$ is extracted using Eq.~(\ref{eq:alphaBN}).}.
For the combination, the $B\to\pi\pi$ and $B\to\rho\rho$ inputs were unchanged, which leads to the $\alpha$ scan plot shown in Fig.~\ref{fig:alphaScan}(b).
It can be seen that with appropriate helicity convention, the $\rho\pi$ curve is flatter around $90^\circ$ and, therefore, its impact on the overall result is much smaller compared to the original case.
The best-fit $\chi^2$ value is 8.4 in the original scenario, and 9.5 in the scenario with aligned conventions. There are $26-9=17$ degrees of freedom.
The statistical significance of the deviation of $\rho\pi$ with respect to the indirect value $91.2^\circ$ is $\sqrt{\Delta\chi^2} = 2.9\sigma$ for the HFLAV public result and $\sqrt{\Delta\chi^2} = 2.1\sigma$, when convention-aligned $U/I$ are used.

The combination of all three modes results in $\alpha = (87.0^{+11.8}_{-4.7})^\circ$, fully consistent with the indirect value $(91.2^{+2.0}_{-1.0})^\circ$~\cite{Charles:2004jd}.
The combined $\chi^2$ improved by 1.4 in the rectified combination.
In addition, as shown in Fig.~\ref{fig:alphaScan}(b), there is a mirror solution at $\alpha$ near $0^\circ$ for both the total combination and when $B\to\pi\pi$ and $B\to\rho\rho$ modes are combined.

\begin{figure}[t]
\centering
\includegraphics[width=\columnwidth]{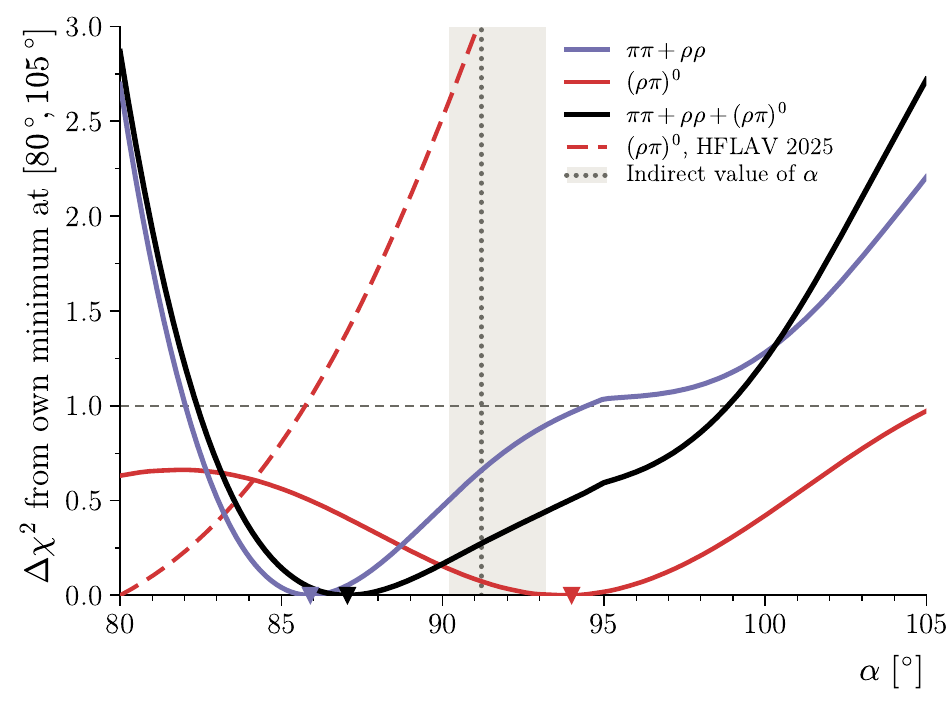}
\caption{One-dimensional $\alpha$ scan of $\chi^2$ for combination of $B\to\pi\pi$ and $B\to\rho\rho$ modes (blue), $B^0\to(\rho\pi)^0$ mode (red), and the total combination of $B\to\pi\pi$, $B\to\rho\rho$, and $B^0\to(\rho\pi)^0$ modes (black). The original $B^0\to(\rho\pi)^0$ contribution as provided by HFLAV~\cite{HeavyFlavorAveragingGroupHFLAV:2024ctg} is plotted as a red dashed line. For clarity, an offset is applied to the plotted curves, so that their minimal value in the $[80^\circ,105^\circ]$ interval is always zero. The indirect value of $\alpha$ with its uncertainty~\cite{Charles:2004jd} is visualized as a vertical gray band.}
\label{fig:alphaScanZoom}
\end{figure}

The larger upper uncertainty is related to the complex structure of the $1-CL$ distributions (Fig.~\ref{fig:alphaScan}); effectively, two $B\to\pi\pi$ peaks contribute, while the $B\to\rho\rho$ solution is in the middle.
For reference, the combination of only $B\to\rho\rho$ and $B\to\pi\pi$ modes leads to $\alpha_{\pi\pi,\rho\rho} = (85.9^{+8.6}_{-3.9})^\circ$. Adding the $B^0\to (\rho\pi)^0$ contribution, which has a shallow local minimum in $\chi^2$ at $94^\circ$, slightly inflates the upper uncertainty of the combination (Fig.~\ref{fig:alphaScanZoom}).
On the other hand, the $2\sigma$ confidence interval slightly shrinks by addition of the $B^0\to(\rho\pi)^0$ mode, from $[79.0^\circ, 109.2^\circ]$ to $[79.1^\circ, 108.1^\circ]$.
In general, compared to the original HFLAV~2025 results, the rectified $B^0\to (\rho\pi)^0$ $\chi^2$ curve is much flatter in the region around the indirect $\alpha$ value and does not push the combination towards lower $\alpha$ values.



In summary, we have demonstrated that in the Snyder-Quinn method, which determines the CKM angle $\alpha$ through time-dependent Dalitz analysis of the $B^0\to \pi^+\pi^-\pi^0$ system, the alignment of the helicity convention used for the $U/I$ parameters measurement and in the subsequent isospin analysis, which determines $\alpha$, plays a crucial role.
All phenomenological studies we have examined, including the original Belle and BaBar analyses and the HFLAV/CKMfitter and UTfit interpretations, exhibit a mismatch between conventions.

Using the appropriate helicity convention reduces the $2.9\sigma$ tension in the $B^0\to(\rho\pi)^0$ mode to $2.1\sigma$ and makes the combined $\alpha$ value, $\alpha = (87.0^{+11.8}_{-4.7})^\circ$, consistent with the indirect determination within uncertainties.

At the current stage, incorporation of the $B^0\to (\rho\pi)^0$ channel does not improve the overall precision of $\alpha$, which is driven by the $B\to\rho\rho$ and $B\to\pi\pi$ modes.
For the precision $B^0\to \pi^+\pi^-\pi^0$ measurement, a careful experimental treatment of the interference regions in the Dalitz plot analysis is essential.
We believe that this contribution together with more advanced analysis techniques will help in future Belle~II and LHCb measurements of this mode~\cite{Belle-II:2018jsg,LHCb:2018roe}.


\bibliography{apssamp}

@article{Lipkin:1991st,
    author = "Lipkin, Harry J. and Nir, Yosef and Quinn, Helen R. and Snyder, A.",
    title = "{Penguin trapping with isospin analysis and \textit{CP} asymmetries in \ensuremath{B} decays}",
    reportNumber = "SLAC-PUB-5445, WIS-91-5-PH",
    doi = "10.1103/PhysRevD.44.1454",
    journal = "Phys. Rev. D",
    volume = "44",
    pages = "1454--1460",
    year = "1991"
}

@article{Snyder:1993mx,
    author = "Snyder, Arthur E. and Quinn, Helen R.",
    title = "{Measuring \textit{CP} asymmetry in $B\to\rho\pi$ decays without ambiguities}",
    reportNumber = "SLAC-PUB-6056",
    doi = "10.1103/PhysRevD.48.2139",
    journal = "Phys. Rev. D",
    volume = "48",
    pages = "2139--2144",
    year = "1993"
}

@article{Charles:2017evz,
    author = "Charles, J. and Deschamps, O. and Descotes-Genon, S. and Niess, V.",
    title = "{Isospin analysis of charmless \ensuremath{B}-meson decays}",
    eprint = "1705.02981",
    archivePrefix = "arXiv",
    primaryClass = "hep-ph",
    doi = "10.1140/epjc/s10052-017-5126-9",
    journal = "Eur. Phys. J. C",
    volume = "77",
    number = "8",
    pages = "574",
    year = "2017"
}

@article{HeavyFlavorAveragingGroupHFLAV:2024ctg,
    author = "Banerjee, Sw. and others",
    collaboration = "Heavy Flavor Averaging Group",
    title = "{Averages of \ensuremath{b}-hadron, \ensuremath{c}-hadron, and {\ensuremath{\tau}}-lepton properties as of 2023}",
    eprint = "2411.18639",
    archivePrefix = "arXiv",
    primaryClass = "hep-ex",
    doi = "10.1103/x87q-tld5",
    journal = "Phys. Rev. D",
    volume = "113",
    number = "1",
    pages = "012008",
    year = "2026"
}

@article{Charles:2004jd,
    author = "Charles, J. and Hocker, Andreas and Lacker, H. and Laplace, S. and Le Diberder, F. R. and Malcles, J. and Ocariz, J. and Pivk, M. and Roos, L.",
    collaboration = "CKMfitter Group",
    title = "{\textit{CP} violation and the CKM matrix: Assessing the impact of the asymmetric $B$ factories}",
    eprint = "hep-ph/0406184",
    archivePrefix = "arXiv",
    reportNumber = "CPT-2004-P-030, LAL-04-21, LAPP-EXP-2004-01, LPNHE-2004-01",
    doi = "10.1140/epjc/s2005-02169-1",
    journal = "Eur. Phys. J. C",
    volume = "41",
    number = "1",
    pages = "1--131",
    year = "2005"
}

@article{BaBar:2013uwm,
    author = "Lees, J. P. and others",
    collaboration = "BaBar Collaboration",
    title = "{Measurement of \textit{CP}-violating asymmetries in $B^0 \to (\rho \pi)^0$ decays using a time-dependent Dalitz plot analysis}",
    eprint = "1304.3503",
    archivePrefix = "arXiv",
    primaryClass = "hep-ex",
    reportNumber = "BABAR-PUB-12-026, SLAC-PUB-15425",
    doi = "10.1103/PhysRevD.88.012003",
    journal = "Phys. Rev. D",
    volume = "88",
    number = "1",
    pages = "012003",
    year = "2013"
}

@article{Belle:2007krm,
    author = "Kusaka, A. and others",
    collaboration = "Belle Collaboration",
    title = "{Measurement of \textit{CP} Asymmetry in a Time-Dependent Dalitz Analysis of \ensuremath{B^0\to (\rho\pi)^0}  and a Constraint on the CKM Angle $\phi_2$}",
    eprint = "hep-ex/0701015",
    archivePrefix = "arXiv",
    reportNumber = "BELLE-PREPRINT-2007-4, KEK-PREPRINT-2006-65",
    doi = "10.1103/PhysRevLett.98.221602",
    journal = "Phys. Rev. Lett.",
    volume = "98",
    pages = "221602",
    year = "2007"
}

@article{Belle:2007jkw,
    author = "Kusaka, A. and others",
    collaboration = "Belle Collaboration",
    title = "{Measurement of \textit{CP} asymmetries and branching fractions in a time-dependent Dalitz analysis of  \ensuremath{B^0\to (\rho\pi)^0}  and a constraint on the quark mixing angle $\phi_2$}",
    eprint = "0710.4974",
    archivePrefix = "arXiv",
    primaryClass = "hep-ex",
    reportNumber = "BELLE-PREPRINT-2007-43, KEK-PREPRINT-2007-47",
    doi = "10.1103/PhysRevD.77.072001",
    journal = "Phys. Rev. D",
    volume = "77",
    pages = "072001",
    year = "2008"
}

@article{UTfit:2022hsi,
    author = "Bona, Marcella and others",
    collaboration = "UTfit group",
    title = "{New UTfit Analysis of the Unitarity Triangle in the Cabibbo-Kobayashi-Maskawa scheme}",
    eprint = "2212.03894",
    archivePrefix = "arXiv",
    primaryClass = "hep-ph",
    reportNumber = "YITP-SB-2022-40",
    doi = "10.1007/s12210-023-01137-5",
    journal = "Rend. Lincei Sci. Fis. Nat.",
    volume = "34",
    pages = "37--57",
    year = "2023"
}

@techreport{LHCb:2018roe,
    author = "Aaij, Roel and others",
    collaboration = "LHCb Collaboration",
    title = "{Physics case for an LHCb Upgrade II - Opportunities in flavour physics, and beyond, in the HL-LHC era}",
    institution = "CERN",
    type = "Report",
    number = "CERN-LHCC-2018-027",
    eprint = "1808.08865",
    archivePrefix = "arXiv",
    primaryClass = "hep-ex",
    year = "2018"
}

@article{Belle-II:2018jsg,
    author = "Altmannshofer, W. and others",
    editor = "Kou, E. and Urquijo, P.",
    collaboration = "Belle~II Collaboration",
    title = "{The Belle II Physics Book}",
    eprint = "1808.10567",
    archivePrefix = "arXiv",
    primaryClass = "hep-ex",
    reportNumber = "KEK Preprint 2018-27, BELLE2-PUB-PH-2018-001, FERMILAB-PUB-18-398-T, JLAB-THY-18-2780, INT-PUB-18-047, UWThPh 2018-26",
    doi = "10.1093/ptep/ptz106",
    journal = "PTEP",
    volume = "2019",
    number = "12",
    pages = "123C01",
    year = "2019",
    note = "[Erratum: PTEP 2020, 029201 (2020)]"
}

@article{Beneke:2003zv,
    author = "Beneke, Martin and Neubert, Matthias",
    title = "{QCD factorization for \ensuremath{B\to PP} and \ensuremath{B\to PV} decays}",
    eprint = "hep-ph/0308039",
    archivePrefix = "arXiv",
    reportNumber = "CLNS-03-1835, PITHA-03-06",
    doi = "10.1016/j.nuclphysb.2003.09.026",
    journal = "Nucl. Phys. B",
    volume = "675",
    pages = "333--415",
    year = "2003"
}

@article{Cabibbo:1963yz,
    author = "Cabibbo, Nicola",
    title = "{Unitary Symmetry and Leptonic Decays}",
    doi = "10.1103/PhysRevLett.10.531",
    journal = "Phys. Rev. Lett.",
    volume = "10",
    pages = "531--533",
    year = "1963"
}

@article{Kobayashi:1973fv,
    author = "Kobayashi, Makoto and Maskawa, Toshihide",
    title = "{\textit{CP} Violation in the Renormalizable Theory of Weak Interaction}",
    reportNumber = "KUNS-242",
    doi = "10.1143/PTP.49.652",
    journal = "Prog. Theor. Phys.",
    volume = "49",
    pages = "652--657",
    year = "1973"
}

@article{Gronau:1990ka,
    author = "Gronau, Michael and London, David",
    title = "{Isospin analysis of \textit{CP} asymmetries in $B$ decays}",
    reportNumber = "DESY-90-106-REV, UDEM-LPN-TH24-REV",
    doi = "10.1103/PhysRevLett.65.3381",
    journal = "Phys. Rev. Lett.",
    volume = "65",
    pages = "3381--3384",
    year = "1990"
}

\end{document}